\documentclass{svjour3a}                     % onecolumn (standard format)
\smartqed  % flush right qed marks, e.g. at end of proof
\usepackage{graphicx}
\usepackage{mathptmx}      % use Times fonts if available on your TeX system
\usepackage{enumerate}
\usepackage{dsfont}

\usepackage[english]{babel}
\usepackage[latin1]{inputenc}
\usepackage{times}
\usepackage{fixmath}
\usepackage{amssymb,amsfonts,amsmath}
\DeclareSymbolFont{extra}{OML}{cmm}{m}{n}
\DeclareMathSymbol{\varrho}{\mathord}{extra}{'045} 
\DeclareMathSymbol{\nu}{\mathord}{extra}{'027}
\DeclareMathSymbol{\zeta}{\mathord}{extra}{'020}
\DeclareMathSymbol{\kappa}{\mathord}{extra}{'024}
\DeclareMathSymbol{\omega}{\mathord}{extra}{'041}
\DeclareMathSymbol{\Ph}{\mathord}{extra}{'10}
\DeclareMathSymbol{\Omega}{\mathord}{extra}{'12}
\DeclareMathSymbol{\alpha}{\mathord}{extra}{'13}
\DeclareMathSymbol{\beta}{\mathord}{extra}{'14}
\DeclareMathSymbol{\gamma}{\mathord}{extra}{'15}
\DeclareMathSymbol{\delta}{\mathord}{extra}{'16}
\DeclareMathSymbol{\eta}{\mathord}{extra}{'21}
\DeclareMathSymbol{\xi}{\mathord}{extra}{'30}
\DeclareMathSymbol{\varepsilon}{\mathord}{extra}{'42}
\DeclareMathSymbol{\varphi}{\mathord}{extra}{'47}
\DeclareMathSymbol{\pi}{\mathord}{extra}{'31}
\DeclareMathSymbol{\phi}{\mathord}{extra}{'36}

\newcommand{\dd}{\mathrm{d}}
\newcommand{\ee}{\mathrm{e}}
\newcommand{\ii}{\mathrm{i}}

\newcommand{\eps}{\varepsilon}

\newcommand{\rW}{\varrho}
\newcommand{\zW}{\zeta}
\newcommand{\A}{\mathcal A}
\newcommand{\Ap}{\mathcal A^+}
\newcommand{\An}{\mathcal A^0}
\newcommand{\Am}{\mathcal A^-}
\newcommand{\CH}{^\textrm{CH}}
\newcommand{\bPhi}{{\mathbold\Phi}}
 \journalname{General Relativity and Gravitation}
\begin{document}

\title{Non-existence of stationary two-black-hole configurations}
%\dedication{WIDMUNG}
\author{Gernot Neugebauer \and J\"org Hennig}


\institute{G. Neugebauer \at
          Theoretisch-Physikalisches Institut,
          Friedrich-Schiller-Universit\"at,
          Max-Wien-Platz 1,\\
          D-07743 Jena, Germany\\
          \email{G.Neugebauer@tpi.uni-jena.de}           %  \\
           \and
           J. Hennig \at
           Max-Planck-Institut f\"ur Gravitationsphysik,
           Albert-Einstein-Institut,
           Am M\"uhlenberg 1,\\
           D-14476 Potsdam, Germany\\
           \email{pjh@aei.mpg.de}           
}

\date{}

\maketitle
%%%%%%%%%%%%%%%%%%%%%%%%%%%%%%%%%%%%%%%%%%%%%%%%%%%%%%%%%%%%%%%%%%%%%%%%%%%
\begin{abstract}
We resume former discussions of the question, whether the spin-spin 
repulsion and the gravitational attraction of two aligned black holes 
can balance each other. To answer the question we formulate a boundary 
value problem for two separate (Killing-) horizons and apply the inverse 
(scattering) method to solve it. Making use of results of Manko, Ruiz 
and Sanabria-G\'omez and a novel black hole criterion, we prove the
non-existence of the equilibrium situation in question.
%\PACS{04.20.Jb \and 04.70.-s \and 04.20.Cv}
\end{abstract}
%%%%%%%%%%%%%%%%%%%%%%%%%%%%%%%%%%%%%%%%%%%%%%%%%%%%%%%%%%%%%%%%%%%%%%%%%%%
\section{Introduction}
\label{sec:intro}
This paper is meant to contribute to the present discussion about the
existence or non-existence of stationary equilibrium configurations
consisting of separate bodies at rest.
Hermann Weyl, whom J\"urgen Ehlers admired especially, was the first
person to consider the problem of two separate static (axisymmetric)
bodies in equilibrium \cite{Weyl}. 
To mention only one modern advancement
in this field we refer to a paper by Beig and
Schoen~\cite{Beig}, who were able to prove a non-existence theorem for a
reflectionally symmetric \emph{static} $n$-body configuration.

Our intention is to involve the interaction of the angular momenta of
rotating bodies (``spin-spin interaction'') which could generate
repulsive effects compensating the omnipresent mass attraction. A
characteristic example for such a stationary configuration could be the
equilibrium between two aligned rotating black holes. We will present
and review a chain of old and new arguments which finally forbid the
 equilibrium situation.

\begin{figure}\centering
 \includegraphics[width=0.34\textwidth]{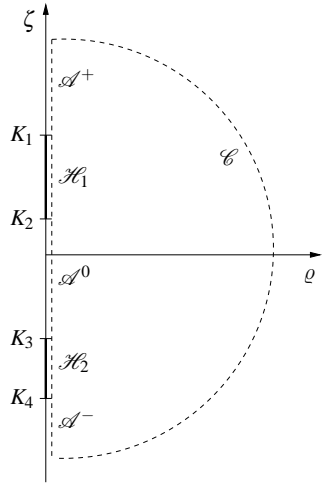}
 \caption{Illustration of the two-black-hole equilibrium
 situation in Weyl coordinates. The event
 horizons $\mathcal H_1$ and $\mathcal H_2$ of the two black holes
 are located in the
 intervals $[K_2,K_1]$ and $[K_4,K_3]$ on the $\zW$-axis, respectively.
 The remaining parts $\mathcal A^\pm$, $\mathcal A^0$
 of the $\zW$-axis correspond to the rotation axis.}
 \label{fig:0}
\end{figure}

Our argumentation is based on a boundary value problem for two separate
(Killing-) horizons (see Fig.~\ref{fig:0}) and the characterization of
\emph{sub-extremal} black holes by Booth and Fairhurst \cite{Booth} and
follows the ideas of Manko and Ruiz \cite{Manko2001} who solved the
equilibrium problem for the so-called double-Kerr solution.

The double-Kerr (more precicely: double-Kerr-NUT)
solution, first derived
in \cite{Neugebauer1980a,Kramer1980},
is a seven parameter solution constructed by a
two-fold B\"acklund transformation of Minkowski space. Since a
single B\"acklund transformation generates the Kerr-NUT solution that
contains, by a special choice of its three parameters, the stationary
black hole solution (Kerr solution) and since B\"acklund
transformations act as a ``nonlinear'' superposition principle, the
double-Kerr-NUT solution was considered to be a good candidate for the
solution of the two horizon problem and extensively discussed in the
literature
\cite{Kramer1980,Kihara1982,Tomimatsu,Hoenselaers1983,Yamazaki,Hoenselaers1984,Dietz,Kramer1986,Manko2000,Krenzer,Manko2001}.
However, there was no argument requiring
that this particular solution be the \emph{only} candidate. In this paper we
will remove this objection and show that the discussion of a boundary
value problem for two separate horizons necessarily leads to the
double-Kerr-NUT solution. Thus we can make use of the equilibrium
conditions for this solution which ensure that the
intervals $\Ap$,
$\An$, $\Am$ (see Fig.~\ref{fig:0}) are regular parts of the axis of
symmetry. After a too restrictive ansatz in \cite{Kramer1980}, 
Tomimatsu and Kihara \mbox{\cite{Kihara1982,Tomimatsu}}
derived and discussed a complete
set of equilibrium
conditions on the axis. Reformulations and numerical studies by
Hoenselaers \cite{Hoenselaers1984} made plausible that
the two gravitational sources (black hole candidates) of the
double-Kerr-NUT solution (located at the intervals $\rW=0$,
$K_1\ge\zW\ge K_2$ and $K_3\ge\zW\ge K_4$) cannot be in equilibrium if
their Komar masses are positive.
The first decisive step toward prove the Hoenselaers conjecture
was taken by Manko, Ruiz and Sanabria-G\'omez \cite{Manko2000},
who derived an explicit
and easily applicable form of the Tomimatsu-Kihara
equilibrium conditions and, as an
important complement, analytical formulae for the Komar masses and
angular momenta of the gravitational sources. Manko and Ruiz
\cite{Manko2001} completed their
non-existence proof by showing that the equilibrium conditions for the
double-Kerr-NUT solution are
indeed violated for positive Komar masses.
 This is, however, a critical point of their analysis.
To the best of our knowledge there is no argument in favour of the
positiveness of the Komar mass. (On the contrary, Ansorg and
Petroff \cite{Ansorg2006} have given a convincing counterexample.)  

In this paper we replace the Komar mass inequality (positivity of the
Komar mass of each black hole) by an inequality connecting angular
momentum and horizon area \cite{Hennig2008a}. This relation is based
on the causal structure of trapped surfaces in the interior vicinity of
the event horizon \cite{Booth}. In this way we can complete the no-go
theorem, avoiding more laborious investigations of the domain outside the
horizons and off the axis of symmetry (e.g., the search for singular rings or
other singularities --- in Sec.~\ref{sing} we will return to that
question). 

%%%%%%%%%%%%%%%%%%%%%%%%%%%%%%%%%%%%%%%%%%%%%%%%%%%%%%%%%%%%%%%%%%%%%%%%%%%
\section{The boundary value problem}
%%%%%%%%%%%%%%%%%%%%%%%%%%%%%%%%%%
\subsection{The boundary conditions\label{sec:2.1}}

The exterior vacuum gravitational field of axisymmetric and stationary
gravitational sources can be described in cylindrical
Weyl-Lewis-Papapetrou coordinates by the line element
\begin{equation}\label{LE}
 \dd s^2 = \ee^{-2U}\big[\ee^{2k}(\dd\rW^2+\dd\zW^2)
           +\rW^2\dd\varphi^2\big]
           -\ee^{2U}(\dd t+a\dd\varphi)^2,
\end{equation}
where the ``Newtonian'' gravitational potential $U$, the
``gravitomagnetic'' potential $a$ and the ``superpotential'' $k$ are
functions of $\rW$ and $\zW$ alone. Figure~\ref{fig:0}
shows the boundaries of the vacuum region: $\Ap$, $\An$, $\Am$ are the
regular parts of the axis of symmetry, $\mathcal H_1$ and
$\mathcal H_2$ are Killing horizons and $\mathcal C$ stands for spatial
infinity. Regularity of the metric along $\Ap$, $\An$, $\Am$ means
elementary flatness and uniqueness on the axis of symmetry,
\begin{equation}\label{eleflat}
 \mathcal A^\pm, \An:\quad a=0,\quad k=0.
\end{equation}
The spacetime has to be flat at large distances from the horizons,
\begin{equation}\label{flat}
 \mathcal C:\quad U\to 0,\quad a\to 0,\quad k\to 0,
\end{equation}
i.e. the line element \eqref{LE} takes a Minkowskian form in cylindrical
space ($\rW,\zW,\varphi$)- time ($t$) coordinates.

The metric \eqref{LE} allows an Abelian group of motions $G_2$ with the
generators (Killing vectors)
\begin{eqnarray}\label{KV}
 &&\xi^i = \delta^i_t,\quad\,\,\,\, \xi^i\xi_i<0\quad\textrm{(stationarity)},\\
 &&\eta^i = \delta^i_\varphi,\quad \eta^i\eta_i>0\quad\textrm{(axisymmetry)},
\end{eqnarray}
where the Kronecker symbols $\delta^i_t$ and $\delta^i_\varphi$ indicate
that $\xi^i$ has only a $t$-component whereas $\eta^i$ points in the
azimuthal $\varphi$-direction along closed circles. Obviously,
\begin{equation}\label{mc}
 \ee^{2U} = -\xi^i\xi_i,\qquad a=-\ee^{-2U}\eta_i\xi^i
\end{equation}
is a coordinate-free representation of the two relativistic
gravitational potentials $U$ and $a$ with the boundary values
\eqref{eleflat}, \eqref{flat}.

In stationary and axisymmetric spacetimes,
the event horizon of a black hole is a Killing horizon which can be
defined by a linear combination $L$ of $\xi$ and $\eta$,
\begin{equation}\label{L}
 L=\xi+\Omega\eta,
\end{equation}
where $\Omega$ is a constant. A connected component of the set of points
with $\ee^{2V}:=-(L,L)=0$, which is a null hypersurface,
$(\dd\ee^{2V},\dd\ee^{2V})=0$, is called a Killing horizon
$\mathcal H(L)$,
\begin{equation}\label{KH}
 \mathcal H(L):\quad \ee^{2V} = -(L,L)=0,\qquad
                     (\dd\ee^{2V},\dd\ee^{2V})=0.
\end{equation}
Since the Lie derivative $\mathcal L_L$ of $\ee^{2V}$ vanishes, we have
$(L,\dd\ee^{2V})=0$. Hence, $L$ and $\ee^{2V}$ being null vectors on
$\mathcal H(L)$ are proportional to each other,
\begin{equation}\label{N}
 \mathcal H(L):\quad \dd\ee^{2V}=-2\kappa L.
\end{equation}
Using the field equations one can show that the \emph{surface gravity}
$\kappa$ is a constant on $\mathcal H(L)$. In Weyl-Lewis-Papapetrou
coordinates the event horizon degenerates to a ``straight line'' and
covers a $\zW$-interval at $\rW=0$ \cite{Carter}.
To formulate boundary conditions on
the horizons $\mathcal H_1$ and $\mathcal H_2$ (see Fig.~\ref{fig:0}) we
make use of \eqref{mc} and \eqref{L} to express $\ee^{2V}$ in terms of
$\ee^{2U}$, $a$ and $\rW$,
\begin{eqnarray}
 \label{BC1}
 \mathcal H_1:\quad\ee^{2V_1}:=\ee^{2U}\left[(1+\Omega_1
 a)^2-\Omega_1^2\rW^2\ee^{-4U}\right]=0,\quad
 \rW=0,\quad K_1\ge\zW\ge K_2,\\
 \label{BC2}
 \mathcal H_2:\quad\ee^{2V_2}:=\ee^{2U}\left[(1+\Omega_2
 a)^2-\Omega_2^2\rW^2\ee^{-4U}\right]=0,\quad
 \rW=0,\quad K_3\ge\zW\ge K_4.
\end{eqnarray}
$\Omega_1$, $\Omega_2$ are the constant angular velocities of the
horizons $\mathcal H_1$, $\mathcal H_2$, respectively.
%%%%%%%%%%%%%%%%%%%%%%%%%%%%%%%%%%
\subsection{The field equations}
The vacuum Einstein equations for the metric potentials $U$, $a$, $k$
are equivalent to the Ernst equation
\begin{equation}\label{Ernst}
 (\Re f)\Big(f_{,\rW\rW}+f_{,\zW\zW} +\frac{1}{\rW}f_{,\rW}\Big)
 = f_{,\rW}^2 + f_{,\zW}^2
\end{equation}
for the complex function
\begin{equation}\label{EP}
 f(\rW,\zW)=\ee^{2U(\rW,\zW)}+\ii b(\rW,\zW),
\end{equation}
where $b$ replaces $a$ via
\begin{equation}\label{a}
 a_{,\rW} = \rW\ee^{-4U}b_{,\zW},\qquad
 a_{,\zW} = -\rW\ee^{-4U}b_{,\rW}
\end{equation}
and $k$ can be calculated from
\begin{eqnarray}
 \label{k1}
 k_{,\rW} & = & \rW\Big[U_{,\rW}^2-U_{,\zW}^2+\frac{1}{4}\ee^{-4U}
         (b_{,\rW}^2-b_{,\zW}^2)\Big],\\
 \label{k2}
 k_{,\zW} & = & 2\rW\Big[U_{,\rW}U_{,\zW}+\frac{1}{4}\ee^{-4U}
         b_{,\rW}b_{,\zW}\Big].
\end{eqnarray}

As a consequence of the Ernst equation \eqref{Ernst}, the integrability
conditions $a_{,\rW\zW}=a_{,\zW\rW}$ and $k_{,\rW\zW}=k_{,\zW\rW}$ are
satisfied such that the metric potentials $a$ and $k$ may be calculated
via line integration from the Ernst potential $f$. Thus the boundary
value problem for the vacuum Einstein equations reduces to a boundary
value problem for the Ernst equation. However, we have to cope with
non-local boundary conditions for the Ernst potential, see
\eqref{eleflat}, \eqref{flat}, \eqref{a}, \eqref{k1}, \eqref{k2}.
Fortunately, these boundary conditions are well-adapted to the ``inverse
method'', which will be applied to solve the boundary value problem.
%%%%%%%%%%%%%%%%%%%%%%%%%%%%%%%%%%
\subsection{The inverse method}

The inverse (scattering) method first applied to solve initial value
problems of special classes of \emph{non-linear} partial differential
equations in many areas of physics (Korteweg-de Vries equation in
hydrodynamics, non-linear Schr\"odinger equation in non-linear optics
etc.) is based on the existence of a linear problem (LP) whose
integrability condition is equivalent to the non-linear differential
equation. Luckily, the Ernst equation has an LP too,
so one can try to tackle boundary value problems for rotating
objects including black holes. We use the LP
\cite{Neugebauer1979,Neugebauer1980b}
\begin{equation}\label{LP}
\begin{aligned}
 \bPhi_{,z} & = \left[\left(\begin{array}{cc}
                  B & 0\\
                  0 & A\end{array}\right)
                  +\lambda\left(\begin{array}{cc}
                  0 & B\\
                  A & 0\end{array}\right)\right]\bPhi,\\
 \bPhi_{,\bar z} & = \left[\left(\begin{array}{cc}
                  \bar A & 0\\
                  0 & \bar B\end{array}\right)
                  +\frac{1}{\lambda}\left(\begin{array}{cc}
                  0 & \bar A\\
                  \bar B & 0\end{array}\right)\right]\bPhi,
\end{aligned}                 
\end{equation}
where $\bPhi(z,\bar z,\lambda)$ is a $2\times2$ matrix depending on the
spectral parameter
\begin{equation}\label{lambda}
\lambda=\sqrt{\frac{K-\ii\bar z}{K+\ii z}}
\end{equation}
as well as on the complex coordinates $z=\rW+\ii\zW$, $\bar
z=\rW-\ii\zW$, whereas
\begin{equation}
 A=\frac{f_{,z}}{f+\bar f},\qquad
 B=\frac{\bar f_{,z}}{f+\bar f}
\end{equation}
and the complex conjugate quantities $\bar A$, $\bar B$ are functions of
$z$, $\bar z$ (or $\rW$, $\zW$) alone and do not depend on the constant
parameter $K$. From the
integrability condition $\bPhi_{,z\bar z}=\bPhi_{,\bar z z}$ and the
relations
\begin{equation}
 \lambda_{,z}=\frac{\lambda}{4\rW}(\lambda^2-1),\qquad
 \lambda_{,\bar z}=\frac{1}{4\rW\lambda}(\lambda^2-1)
\end{equation}
it follows that the $\lambda$-independent coefficients of a matrix
polynomial in $\lambda$ have to vanish. The result is the Ernst equation
\eqref{Ernst}. Vice versa, the matrix $\bPhi$ calculated from \eqref{LP}
does not depend on the path of integration if $f$ is a solution to the
Ernst equation. The idea of the inverse (scattering) method is to
construct $\bPhi$, for fixed but arbitrary values of $z$, $\bar z$, as a
holomorphic function of $\lambda$ and to calculate $f(\rW,\zW)$ from
$\bPhi$. To obtain the dependence on $\lambda$ for the two-horizon
problem we have to integrate the linear system \eqref{LP} along the
dashed line $\mathcal D=\Am\cup\mathcal H_2\cup\An\cup\mathcal H_1\cup\Ap\cup
\mathcal C$ making use of the boundary conditions \eqref{eleflat},
\eqref{flat}, \eqref{BC1}, \eqref{BC2}.
As was shown in \cite{Neugebauer2000} and \cite{Neugebauer2003}
(see Eqs. (34), (45), (57), (58) in \cite{Neugebauer2003})
the result of the integration is a matrix representation of the axis
values of the Ernst potential
$f^+(\zW)\equiv f^+(\rW=0,\zW)=\ee^{2U^+(\zW)}+\ii b^+(\zW)$
on $\Ap$ in terms of the parameters $K_i$
($i=1,\dots,4$), $f_i=f(\rW=0, \zW=K_i)$ and the angular velocities
$\Omega_1=\Omega^{(1)}=\Omega^{(2)}$,
$\Omega_2=\Omega^{(3)}=\Omega^{(4)}$:
\begin{equation}\label{Nmat}
 \mathcal N = \prod\limits_{n=1}^4\left({\bf 1}
               +\frac{{\bf F}_n}{2\ii\Omega^{(n)}(\zW-K_n)}\right),
\end{equation}
where
\begin{equation}\label{Nmat1}
 \mathcal N := \ee^{-2U^+(\zW)}\left(\begin{array}{cc}
               1 & -\ii b^+(\zW)\\[1ex]
               \ii b^+(\zW) & \  f^+(\zW)\bar f^+(\zW)
               \end{array}\right)
\end{equation}
and
\begin{equation}
 {\bf F}_n :=(-1)^n\left(\begin{array}{cc}
             f_n & -1\\[1ex]
             f_n^2 & \  -f_n
             \end{array}\right).
\end{equation}
Obviously, the sum of the off-diagonal elements of $\mathcal N$ has to
vanish, $\mathcal N_{12}+\mathcal N_{21}=0$, whence
\begin{equation}
 \mathrm{tr} \left[\left(\begin{array}{cc}
            0 & 1\\ 1 & 0\end{array}\right)
             \prod\limits_{n=1}^4\left(
             {\bf 1}+\frac{{\bf F}_n}{2\ii\Omega^{(n)}(\zW-K_n)}\right)
             \right] = 0.
\end{equation}
Since this equation holds identically in $\zW$, one obtains four
constraints among $\Omega_1$, $\Omega_2$; $K_3-K_4$, $K_2-K_3$,
$K_1-K_2$; $f_1$,\dots $f_4$. Particularly, the $1/\zW$-term yields
\begin{equation}\label{omratio}
 \frac{\Omega_1}{\Omega_2}=\frac{f_1^2-f_2^2}{f_4^2-f_3^2}.
\end{equation}
Hence the three remaining constraints enable us to express the three
similarity variables $\Omega_1(K_3-K_4)$, $\Omega_1(K_1-K_2)$,
$\Omega_1(K_2-K_3)$, or, alternatively, $\Omega_2(K_3-K_4)$,
$\Omega_2(K_1-K_2)$,\linebreak
$\Omega_2(K_2-K_3)$, in terms of the four values of
the Ernst potential $f_1,\dots,f_4$. It turns out that the
constancy of the surface gravity, or, alternatively, the condition $k=0$
on $\mathcal A^\pm$, $\mathcal A^0$, gives rise 
to one further constraint. Since the axis
values $f^+(\zW)$ determine the solution of the Ernst equation uniquely,
one may expect a three (real) parameter solution of the two-horizon
problem (see remark on Eq. \eqref{dimless}).
%%%%%%%%%%%%%%%%%%%%%%%%%%%%%%%%%%
\subsection{The double-Kerr-NUT solution}

According to \eqref{Nmat} and \eqref{Nmat1},
the axis potential $f^+(\zW)$ is a quotient of
two polynomials in $\zW$. To determine the degree of the polynomials we
compare the polynomial structure of the matrix elements \eqref{Nmat1},
\begin{eqnarray}
 \label{e2U}
 \ee^{2U^+(\zW)} & = &
 \frac{(\zW-K_1)(\zW-K_2)(\zW-K_3)(\zW-K_4)}{p_4(\zW)},\\
 \label{b}
 \ii b^+(\zW) & = & \ii\frac{p_2(\zW)}{p_4(\zW)},\\
 \label{ffq}
 f^+(\zW)\overline{f^+(\zW)} & =  & \frac{\pi_4(\zW)}{p_4(\zW)},
\end{eqnarray}
where $p_4$, $\pi_4$ are real normalized polynomials of the fourth
degree (the fourth power coefficient is equal to one) and the real
polynomial $p_2(\zW)$ is of second degree due to
\eqref{omratio}. Replacing $f^+$ and $\bar f^+$ in the third
equation by the combination $f^+=\ee^{2U^+}+\ii b^+$ and its complex
conjugate from the first and the second equation, we get the condition
\begin{eqnarray}\label{cond}
 &&\left[(\zW-K_1)(\zW-K_2)(\zW-K_3)(\zW-K_4)+\ii p_2\right]\times\nonumber\\
 &&\quad\times\left[(\zW-K_1)(\zW-K_2)(\zW-K_3)(\zW-K_4)-\ii p_2\right]
   = \pi_4(\zW)p_4(\zW).
\end{eqnarray}
Identifying the zeros of both sides, we see that each bracket of the left
hand side has to have two zeros of $\pi_4$ as well as of $p_4$ (note
that the brackets must be complex conjugate to each other). Therefore,
$f^+(\zW)$ has to be the quotient of two polynomials of second degree,
\begin{equation}\label{av}
 f^+(\zW)=\frac{n_2(\zW)}{d_2(\zW)},
\end{equation}
where the numerator polynomial and the denominator polynomial have the
structure
\begin{equation}
\begin{aligned}
 n_2(\zW)& = \zW^2+b\zW+a,\\
 d_2(\zW)& = \zW^2+e\zW+d,
\end{aligned}
\end{equation}
and $a$, $b$, $d$, $e$ are complex constants.

Preparing the continuation of $f^+(\zW)$ off the axis of symmetry into
the $\rW$-$\zW$ plane (see Fig.~\ref{fig:0}) we replace these constants
by the
appropriate parameters $\bar d_2(K_i)/d_2(K_i)$ and\linebreak
$\bar n_2(K_i)/n_2(K_i)$
,
\begin{equation}
 \alpha_i = \frac{K_i^2+\bar e K_i +\bar d}{K_i^2+eK_i+d},\qquad
 \beta_i = \frac{K_i^2+\bar b K_i +\bar a}{K_i^2+bK_i+a},\qquad
 \alpha_i\bar\alpha_i=1,\qquad \beta_i\bar\beta_i=1.
\end{equation}
Equation \eqref{e2U} implies $\ee^{2U^+(K_i)}=0$ and therefore
\begin{equation}
 \overline{f^+(K_i)}=-f^+(K_i),\qquad i=1,\dots,4,
\end{equation}
whence
\begin{equation}\label{beal}
 \beta_i=-\alpha_i.
\end{equation}
In a next step we solve the linear algebraic system
\begin{equation}
 \bar e K_i +\bar d - e\alpha_i K_i - d\alpha_i
 = K_i^2(\alpha_i-1),\qquad i=1,\dots,4,
\end{equation}
to obtain $e$, $d$ and finally $d_2(\zW)$ in terms of $\alpha_i$, $K_i$
($i=1,\dots,4$). Because of \eqref{beal} $n_2(\zW)$ can simply be read
off from $d_2(\zW)$ by replacing $\alpha_i$ by $-\alpha_i$
($i=1,\dots,4$). Thus we arrive at the determinant representation
\begin{equation}\label{axisf}
 f^+(\zW) = \frac{\left|\begin{array}{ccccc}
   1 & K_1^2  & K_2^2  & K_3^2 & K_4^2\\[1ex]
  \ 1\ & \ \alpha_1K_1(\zW-K_1)\ & \ \alpha_2K_2(\zW-K_2)\ &
   \ \alpha_3K_3(\zW-K_3)\ & \ \alpha_4K_4(\zW-K_4)\ \\[1ex]
   0 & K_1    & K_2    & K_3   & K_4\\[1ex]
   0 & \alpha_1(\zW-K_1) & \alpha_2(\zW-K_2) &
   \alpha_3(\zW-K_3) & \alpha_4(\zW-K_4)\\[1ex]
   0 & 1      & 1      & 1     & 1             
   \end{array}\right|}
   {\left|\begin{array}{ccccc}
   1 & K_1^2  & K_2^2  & K_3^2 & K_4^2\\[1ex]
  -1 & \ \alpha_1K_1(\zW-K_1)\ & \ \alpha_2K_2(\zW-K_2)\ &
   \ \alpha_3K_3(\zW-K_3)\ & \ \alpha_4K_4(\zW-K_4)\ \\[1ex]
   0 & K_1    & K_2    & K_3   & K_4\\[1ex]
   0 & \alpha_1(\zW-K_1) & \alpha_2(\zW-K_2) &
   \alpha_3(\zW-K_3) & \alpha_4(\zW-K_4)\\[1ex]
   0 & 1      & 1      & 1     & 1             
   \end{array}\right|}   
\end{equation}
for the Ernst potential $f(\zW)$ on the axis $\Ap$.

We will now construct $f(\rW, \zW)$. It can be shown that the axis
values on $\Ap$ determine the Ernst potential $f(\rW,\zW)$ everywhere in
the $\rW$-$\zW$ plane. Hence, if we find a continuation $f(\rW,\zW)$ of
$f^+(\zW)$ for all $\rW\ge0$ and can prove that it satisfies the Ernst
equation \eqref{Ernst} we have achieved our goal. Introducing the
``distances'' $r_i$ from the points $\rW=0$, $\zW=K_i$ by
\begin{equation}\label{ri}
 r_i := \sqrt{(\zW-K_i)^2+\rW^2}\ge0,\qquad i=1,\dots,4,
\end{equation}
with the property
\begin{equation}
 \Ap:\quad r_i=\zW-K_i,\qquad i=1,\dots,4,
\end{equation}
and replacing the expressions $\zW-K_i$ ($i=1,\dots,4$) in \eqref{axisf}
by $r_i$ we arrive at
\begin{equation}\label{genf}
 f(\rW,\zW) = \frac{\left|\begin{array}{ccccc}
   1 & K_1^2  & K_2^2  & K_3^2 & K_4^2\\[1ex]
  \ 1\ & \ \alpha_1K_1r_1\ & \ \alpha_2K_2r_2\ &
   \ \alpha_3K_3r_3\ & \ \alpha_4K_4r_4\ \\[1ex]
   0 & K_1    & K_2    & K_3   & K_4\\[1ex]
   0 & \alpha_1r_1 & \alpha_2r_2 &
   \alpha_3r_3 & \alpha_4r_4\\[1ex]
   0 & 1      & 1      & 1     & 1             
   \end{array}\right|}
   {\left|\begin{array}{ccccc}
   1 & K_1^2  & K_2^2  & K_3^2 & K_4^2\\[1ex]
  -1 & \ \alpha_1K_1r_1\ & \ \alpha_2K_2r_2\ &
   \ \alpha_3K_3r_3\ & \ \alpha_4K_4r_4\ \\[1ex]
   0 & K_1    & K_2    & K_3   & K_4\\[1ex]
   0 & \alpha_1r_1 & \alpha_2r_2 &
   \alpha_3r_3 & \alpha_4r_4\\[1ex]
   0 & 1      & 1      & 1     & 1             
   \end{array}\right|}.   
\end{equation}
A straightforward calculation shows that $f$, as defined in
\eqref{genf}, is indeed a solution of the Ernst equation
\eqref{Ernst}\footnote{The procedure may seem rather tricky. In fact
it reflects steps of the inverse scattering method whose explanation is
outside the scope of this paper.}.
As we have already mentioned, the remaining gravitational potentials
$k$, $a$ ($\ee^{2U}=\Re f$!) can be calculated from $f$ via line
integrals. This solution of the vacuum Einstein equations represented by
the Ernst potential $f(\rW,\zW)$ with the axis values \eqref{av},
\eqref{axisf} is known as the double-Kerr-NUT
solution. It depends on seven real parameters: four real arguments of
$\alpha_i$, $\alpha_i\bar\alpha_i=1$ ($i=1,\dots,4$) plus three
differences $K_1-K_2$, $K_3-K_4$ (``length'' of horizons), $K_2-K_3$
(``distance'' between the horizons). (Note that the configuration as
sketched in Fig.~\ref{fig:0} can be translated along the $\zW$-axis.)
Hence, \emph{the solution of the two-horizon problem is a (particular)
double-Kerr-NUT solution.}
%%%%%%%%%%%%%%%%%%%%%%%%%%%%%%%%%%
\subsection{The equilibrium conditions}
The double-Kerr-NUT solution in the form \eqref{genf} was presented and
discussed in \cite{Kramer1980} as a particular ($N=2$) case of the
$N$-soliton solution \cite{Neugebauer1980a,Neugebauer1980b}\footnote{A
misprint in Ref. \cite{Neugebauer1980a} was corrected at the end of
Ref. \cite{Neugebauer1980b}.} of the Ernst
equation generated by the application of $N$ B\"acklund transformations
to an arbitrary seed solution. Applying the boundary conditions
\eqref{eleflat}, \eqref{flat} to the representation \eqref{genf},
Tomimatsu and Kihara \cite{Tomimatsu} derived a complete
set of algebraic equilibrium conditions on the axis of symmetry
between the parameters
$\alpha_i$, $K_i$ ($i=1,\dots,4$). Particular solutions of the algebraic
system involving numerical results were discussed by Hoenselaers
\cite{Hoenselaers1984}, who came to conjecture
that the double-Kerr-NUT solution cannot describe equilibrium
between two aligned rotating black holes with positive Komar masses. 
Hoenselaers and Dietz \cite{Dietz,Hoenselaers1983} and Krenzer
\cite{Krenzer} were able to prove this conjecture for symmetric
configurations ($K_1-K_2=K_3-K_4$, $\Omega_1=\Omega_2$).

The final explicit solution of the Tomimatsu-Kihara equilibrium
conditions was found by Manko, Ruiz and Sanabria-G\'omez
\cite{Manko2000}. Following their idea, we start with the condition
$k=0$ on $\mathcal A^\pm$, $\An$,
\begin{equation}
 \mathcal A^\pm, \An:\quad k=0
\end{equation}
and apply it to $k$ calculated from $f$, see, e.g. \cite{Kramer1986}. The
only condition is
\begin{equation}\label{cond1}
 \alpha_1\alpha_2+\alpha_3\alpha_4=0.
\end{equation}
Combining this result with the two conditions derived from $a=0$ on
$\mathcal A^\pm$, $\An$ ($a$ again calculated from $f$, see,
e.g. \cite{Kramer1986}) one obtains
\begin{equation}
 \begin{aligned}\label{alpha}
  \frac{(1-\alpha_4)^2}{\alpha_4}\gamma &= \frac{(1-\alpha_3)^2}{\alpha_3},
  \qquad
   \gamma:=\frac{K_{14}K_{24}}{K_{13}K_{23}},\\
  \frac{(1+\alpha_2)^2}{\alpha_2}\gamma' &= \frac{(1+\alpha_1)^2}{\alpha_1},
  \qquad
   \gamma':=\frac{K_{23}K_{24}}{K_{13}K_{14}}, 
 \end{aligned}
\end{equation}
where
\begin{equation}
 K_{ij}:=K_i-K_j,\qquad i,j=1,\dots,4.
\end{equation}
Introducing the relative horizon ``lengths''
\begin{equation}
 l_1=\frac{K_{12}}{K_{23}},\qquad l_2=\frac{K_{34}}{K_{23}}
\end{equation}
we may express $\gamma$, $\gamma'$ by the scaled quantities $l_1$, $l_2$
alone,
\begin{equation}
 \gamma = \frac{(1+l_2)(1+l_1+l_2)}{1+l_1},\qquad
 \gamma'= \frac{1+l_2}{(1+l_1)(1+l_1+l_2)}.
\end{equation}
Setting
\begin{equation}
 \alpha_3\alpha_4 = -\alpha_1\alpha_2 \equiv \alpha^2\qquad
 (\alpha\bar\alpha = 1)
\end{equation}
to satisfy \eqref{cond1} one obtains the $\alpha_i$ ($i=1,\dots,4$) from
\eqref{alpha} in terms of the three real parameters $\gamma$, $\gamma'$
(or, alternatively, $l_1$, $l_2$) and $\arg\alpha=\phi$
($\alpha=\ee^{\ii\phi}$) \cite{Manko2000}
\begin{equation}\label{alpha1}
 \begin{aligned}
  \alpha_1 &= \frac{w'\alpha^2+\ii\eps\alpha}{w'-\ii\eps\alpha},\qquad
  \alpha_2 = \frac{\alpha^2+\ii w'\eps\alpha}{1-\ii w'\eps\alpha},\\
  \alpha_3 &= \frac{w\alpha^2-\alpha}{w-\alpha},\qquad
  \alpha_4  = \frac{\alpha^2-w\alpha}{1-w\alpha},
 \end{aligned}
\end{equation}
where
\begin{equation}
 w':=|\sqrt{\gamma'}|\in(0,1],\qquad
 w:=|\sqrt{\gamma}|\in[1,\infty),\qquad
 \eps=\pm1.
\end{equation}
Here $l_1$, $l_2$ are arbitrary positive constants and $\alpha$ is a
periodic function of $\phi$, $\alpha=\ee^{\ii\phi}$. 

With the aid of the relations \eqref{alpha1} Manko and Ruiz
\cite{Manko2001} were able to calculate the Komar masses $M_1$, $M_2$ belonging
to the horizons $\mathcal H_1$, $\mathcal H_2$, respectively, and show
that positive Komar masses are incompatible with the equilibrium conditions.

A concise reformulation of the double-Kerr-NUT solution \eqref{genf} was
derived by Yamazaki \cite{Yamazaki},
\begin{equation}\label{Yamazaki}
 f(\rW,\zW) = \frac{\left|\begin{array}{cc}
              R_{12}-1 & R_{14}-1\\
              R_{23}-1 & R_{34}-1\end{array}\right|}
              {\left|\begin{array}{cc}
              R_{12}+1 & R_{14}+1\\
              R_{23}+1 & R_{34}+1\end{array}\right|},\qquad
 R_{ij}:=\frac{\alpha_ir_i-\alpha_jr_j}{K_{ij}},
\end{equation}
whereby the $\alpha_i$, ($i=1,\dots,4$) have to be taken from
\eqref{alpha1}. Obviously, one can introduce dimensionless coordinates
$\tilde\rW$, $\tilde\zW$ via
\begin{equation}\label{dimless}
 \tilde\rW = \frac{\rW}{K_{23}},\qquad
 \tilde\zW = \frac{\zW-K_1}{K_{23}}
\end{equation}
and see directly that the Ernst potential, as a function of $\tilde\rW$ and
$\tilde\zW$, depends only on the three parameters $l_1$, $l_2$, $\phi$.
We will make use of the formulation \eqref{Yamazaki}
in the subsequent sections.

%%%%%%%%%%%%%%%%%%%%%%%%%%%%%%%%%%%%%%%%%%%%%%%%%%%%%%%%%%%%%%%%%%%%%%%%%%%
\section{Thermodynamics of the two-horizon solution}

%%%%%%%%%%%%%%%%%%%%%%%%%%%%%%%%%%
\subsection{Thermodynamic quantities}

The best way to get a systematic survey of the relevant physical
parameters (state variables) of a two-black-hole system and relations
among them is to resort to the framework of black hole
thermodynamics. This theory tells us that the total mass $M$ of the
system is a thermodynamic potential expressed in terms of
the independent extensive
quantities: horizon areas $A_1$, $A_2$ and angular momenta $J_1$,
$J_2$ of the two black holes. As a consequence of the Gibbs formula (see
Eq.~\eqref{Gibbs} below), the intensive state variables angular
velocities $\Omega_1$, $\Omega_2$ and surface gravities $\kappa_1$,
$\kappa_2$ are functions of the independent quantities
too. Furthermore, the individual Komar masses $M_1$, $M_2$ could play a
role. It turns
out that all quantities can be calculated from the Ernst potential and
its  derivatives in the points of intersection of horizon and symmetry
axis ($\rW=0$, $\zW=K_i$, $i=1,\dots,4$).

By integrating parts of the Einstein equations over the two horizons $\mathcal H_1$, $\mathcal H_2$ we obtain the following relations, 
\begin{equation}\label{area}
 \kappa_1 A_1 = 2\pi(K_1-K_2),\qquad
 \kappa_2 A_2 = 2\pi(K_3-K_4),
\end{equation}
\begin{equation}\label{mass}
 \Omega_1 M_1 = \frac{\ii}{4}(f_1-f_2),\qquad
 \Omega_2 M_2 = \frac{\ii}{4}(f_3-f_4),
\end{equation}
\begin{equation}\label{momentum}
 \Omega_1 J_1 = \frac{M_1}{2}-\frac{1}{4}(K_1-K_2),\qquad
 \Omega_2 J_2 = \frac{M_2}{2}-\frac{1}{4}(K_3-K_4),
\end{equation}
where $f_i = f(\rW=0,\zW=K_i)$.
Starting with the properties of the Killing vector $L=\xi+\Omega\eta$ on
the horizon $\mathcal H$ (see Sec.~\ref{sec:2.1}) one can show that
\begin{equation}\label{intensive}
 \kappa_1+\ii\Omega_1 = \frac{1}{2}f^+_{,\zW}|_{\zW=K_1},\qquad
 \kappa_2+\ii\Omega_2 = \frac{1}{2}f^0_{,\zW}|_{\zW=K_3},
\end{equation}
where $f^+$ and $f^0$ are the axis potentials on $\A^+$ and $\A^0$,
respectively.
A direct consequence of \eqref{area} and \eqref{momentum} are the
Smarr formulae \cite{Smarr} 
\begin{equation}\label{Smarr1}
   M_i=2\Omega_i J_i 
         +\frac{\kappa_i}{4\pi} A_i,\qquad i=1,2.
\end{equation}

In order to calculate the ADM mass $M$, one can make use of
the asymptotic behaviour
\begin{equation}
 f=1-\frac{2M}{r}\quad\textrm{for}\quad r\to\infty,
\end{equation}
where $r^2=\rW^2+\zW^2$. Evaluation on
$\Ap$ leads to
\begin{equation}
 M=\frac{1}{2}\lim\limits_{\zW\to\infty}(1-f^+)\zW.
\end{equation}
Interestingly, the explicit calculation shows that
\begin{equation}
 M=M_1 + M_2
\end{equation}
holds for the $3$-parameter solution,
i.e. possibly present space-time singularities %\linebreak
(see Sec.~\ref{sing}) do not contribute to the ADM mass $M$.
As a consequence, we obtain the Smarr
formula
\begin{equation}\label{Smarr}
   M=\sum\limits_{i=1}^2
          \left(2\Omega_i J_i 
         +\frac{\kappa_i}{4\pi} A_i\right)
\end{equation}
for the total mass $M$.

%%%%%%%%%%%%%%%%%%%%%%%%%%%%%%%%%%
\subsection{Gibbs formula}
A regular axisymmetric and stationary vacuum spacetime with black holes
obeys the Gibbs formula (``first law of black hole thermodynamics'')
\begin{equation}\label{Gibbs}
  \delta M=\sum\limits_i
          \left(\Omega_i\delta J_i 
         +\frac{\kappa_i}{8\pi}\delta A_i\right),
\end{equation}
see \cite{BCH}. Hence, the total ADM mass $M$ of the spacetime,
as a function of the extensive quantities $J_i$ and $A_i$,
\begin{equation}
 M=M(J_i, A_i),
\end{equation}
is a thermodynamic potential and infinitesimal mass changes $\delta M$
between neighbouring solutions are given by \eqref{Gibbs}. 

There could be spacetime singularities outside the event horizon (see
Sec.~\ref{sing}). Hence it is not clear
\emph{a priori} whether the Gibbs formula
\eqref{Gibbs} also holds for the special double-Kerr-NUT solution
\eqref{Yamazaki}, \eqref{alpha1}.
Therefore one has to test the validity of \eqref{Gibbs} from the outset.  
For that purpose, we use the formulae from the previous subsection and
the Ernst potential \eqref{Yamazaki}, \eqref{alpha1}
to obtain expressions for 
$M$, $J_1$, $J_2$, $\Omega_1$, $\Omega_2$, $\kappa_1$, $\kappa_2$,
$A_1$ and $A_2$. Obviously, all these quantities can be written in
terms of the four parameters
\begin{equation}
  (P_1,\dots,P_4)=(K_{23}\equiv K_2-K_3, w, w', \alpha).
\end{equation}
As an example, the total mass has the explicit form
\begin{equation}
 M=-\frac{K_2-K_3}{2}\left(1+\frac{w}{w'}\right)
          \frac{\tilde M}
          {\tilde M+\eps\sin\phi\cos\phi}
\end{equation}
with
\begin{equation}
 \tilde M:=1+\frac{\eps}{2}\left(w'+\frac{1}{w'}\right)\sin\phi
            -\frac{1}{2}\left(w+\frac{1}{w}\right)\cos\phi.
\end{equation}
Equation~\eqref{Gibbs} is equivalent to the four equations
\begin{equation}\label{Gibbs2}
    \frac{\partial M}{\partial P_i} =
          \Omega_1 \frac{\partial J_1}{\partial P_i}
         +\Omega_2 \frac{\partial J_2}{\partial P_i}
         +\frac{\kappa_1}{8\pi}\frac{\partial A_1}{\partial P_i}
         +\frac{\kappa_2}{8\pi}\frac{\partial A_2}{\partial P_i},\quad
         i=1,\dots,4.
\end{equation}
A straightforward  calculation shows that \eqref{Gibbs2} is indeed satisfied.
Therefore, we may conclude that possible singularities
\emph{do not} contribute to $\delta M$ and the first law of thermodynamics
\eqref{Gibbs} holds.
%%%%%%%%%%%%%%%%%%%%%%%%%%%%%%%%%%%%%%%%%%%%%%%%%%%%%%%%%%%%%%%%%%%%%%%%%%%
\section{The sub-extremality of black holes\label{sub}}

Following Booth and Fairhurst \cite{Booth}, we will assume that
a physically reasonable non-degener\-ate\footnote{The degenerate
(extremal) case requires special attention.} black hole
should be \emph{sub-extremal}, i.e. 
characterized through the existence of trapped surfaces (surfaces
with a negative expansion of outgoing null geodesics) in every
sufficiently small interior neighbourhood of the event horizon. 
It can be shown \cite{Hennig2008a} that any such axisymmetric and
stationary sub-extremal
black hole satisfies the inequality\footnote{Note that the
inequality \eqref{inequality} can be generalized to
the Einstein-Maxwell case, i.e. to electrically charged black holes, see
\cite{Hennig2008b}.}
\begin{equation}\label{inequality}
 8\pi|J|<A,
\end{equation}
i.e. for given event horizon area $A$, there exists an upper bound for
the absolute value of the angular momentum $|J|$.

In order to test explicitly
whether the two
gravitational sources with the horizons $\mathcal H_1$, $\mathcal H_2$
(tentative black holes) 
in the double-Kerr-NUT solution \eqref{Yamazaki}, \eqref{alpha1}
satisfy this inequality, we calculate the quantities
\begin{equation}
p_i:=\frac{8\pi J_i}{A_i},\qquad i=1,2.
\end{equation}
We obtain the remarkably simple expressions
\begin{equation}\label{pJ}
 p_1 = \eps\frac{1+\Ph w'}{w'(\Ph+w')},\qquad
 p_2 = \eps\frac{w(w-\Ph)}{1-w\Ph},
\end{equation}
where
\begin{equation}
 \Ph:=\cos\phi+\eps\sin\phi\in[-\sqrt{2},\sqrt{2}].
\end{equation}
Hence, the inequality \eqref{inequality} for each of the two black
holes is equivalent to
\begin{equation}
  p_1^2-1 \equiv (1-w'^2)\frac{w'^2+2\Ph w'+1}{w'^2(\Ph+w')^2}<0
\end{equation}
and
\begin{equation}
  p_2^2-1 \equiv (w^2-1)\frac{w^2-2\Ph w+1}{(w\Ph-1)^2}<0,
\end{equation}
respectively.
Taking into account the allowed parameter ranges $w\in[1,\infty)$,
$w'\in(0,1]$, these inequalities can only hold if
\begin{equation}
w'^2+2\Ph w'+1<0\qquad\textrm{and}\qquad
w^2-2\Ph w+1<0.
\end{equation}
However, this implies $\Ph w'<0$ and $\Ph w>0$ in
contradiction to $w'>0$ and $w>0$.

Thus \emph{we have proved the non-existence of a stationary and axisymmetric
two-black-hole configuration with separate horizons} (see
Fig.~\ref{fig:0}), i.e. the spin-spin repulsion of two aligned black
holes cannot compensate for their gravitational attraction.
The non-existence theorem is essentially based on the
inequality \eqref{inequality} that is as shown in \cite{Hennig2008a} a
consequence of a defining geometrical black hole property
\cite{Booth} (the case of extremal black holes requires special
attention).

%%%%%%%%%%%%%%%%%%%%%%%%%%%%%%%%%%%%%%%%%%%%%%%%%%%%%%%%%%%%%%%%%%%%%%%%%%%
\section{Further properties of the double-Kerr-NUT solution}

As we have seen in the previous section, the equilibrium of two aligned
black holes is impossible. The only candidate for a solution of the
balance problem --- the double-Kerr-NUT solution --- has to be dismissed
as a physically irrelevant solution as discussed above. Nevertheless, it
is interesting to study further properties of the solution
\eqref{Yamazaki}, \eqref{alpha1}. In the
following two subsections we comment shortly on the interior
of the two gravitational sources and give numerical evidence for the
existence of singularities outside the horizons
$\mathcal H_1$ and $\mathcal H_2$. 

%%%%%%%%%%%%%%%%%%%%%%%%%%%%%%%%%%
\subsection{The interior of black holes}

It was shown in \cite{Ansorg2008} that every axisymmetric and stationary
black hole,
which is regular in an \emph{exterior} neighbourhood of the event
horizon, also possesses a regular \emph{interior} region inside
the event horizon. In particular, there always exists a regular inner Cauchy
horizon and the inner solution does not develop singularities before
this horizon is reached. Moreover, the spacetime is even regular at the
Cauchy horizon, provided that the angular momentum $J$ of the black hole
does not vanish. Remarkably, the areas $A$ and $A\CH$ of event and
inner Cauchy horizon satisfy the equation\footnote{Note that these
statements can be
generalized to black hole spacetimes with electromagnetic
fields, see \cite{Ansorg2009,Hennig2009}.} 
\begin{equation}\label{equality}
 (8\pi J)^2 =A\CH A.
\end{equation}

It is interesting to test this relation explicitly for both of the two
gravitational sources in the double-Kerr-NUT solution \eqref{Yamazaki},
\eqref{alpha1}. For that purpose, we
calculate
the areas $\mathcal A_1$, $\mathcal A_2$ of the horizons $\mathcal H_1$,
$\mathcal H_2$ using \eqref{area} and \eqref{intensive}.
As shown in \cite{Ansorg2008}, the areas of Cauchy horizons can be
calculated from analytical continuations
of $f^+$ and $f^0$ into regions with $\zW<K_1$ and
$\zW<K_3$, respectively. In this way we obtain
\begin{equation}\label{A2}
 A\CH_1=-4\pi\frac{K_1-K_2}{\Re f^+_{,\zW}|_{\zW=K_2}},\qquad
 A\CH_2=-4\pi\frac{K_3-K_4}{\Re f^0_{,\zW}|_{\zW=K_4}}.
\end{equation}
Using these formulae, the explicit calculation shows that the equations
\begin{equation}\label{equality1}
 (8\pi J_i)^2=A_i\CH A_i,\qquad i=1,2,
\end{equation}
are indeed satisfied, i.e. \eqref{equality} holds
for both gravitational sources.
%%%%%%%%%%%%%%%%%%%%%%%%%%%%%%%%%%
\subsection{Singularities outside the black holes\label{sing}}

As we have proved in Sec.~\ref{sub}, at least one of the two ``black holes''
in the double-Kerr-NUT solution \eqref{Yamazaki}, \eqref{alpha1}
is not sub-extremal for which reason the
solution is not physically reasonable.
It may then well be that singularities outside the horizons appear.

\begin{figure}
 \includegraphics[width=\textwidth]{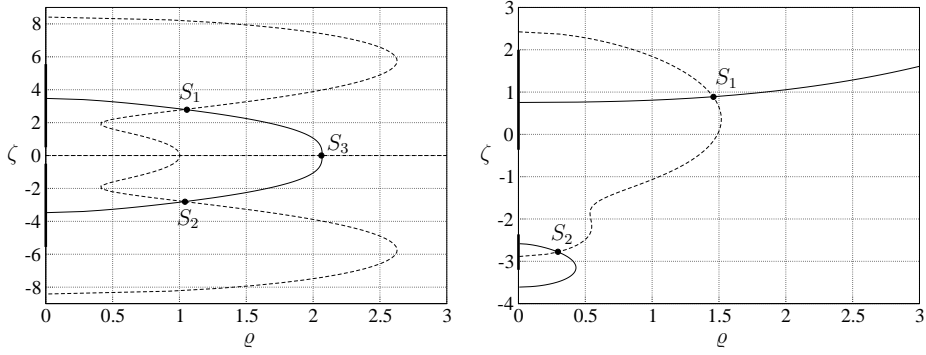}
 \caption{Singularities of the Ernst potential: The plots show, for two
 different configurations, curves
 along which the real part (solid curve) and imaginary part
 (dashed curve) of the determinant in the denominator of the Ernst
 potential $f$ vanish.
 At the intersection
 points $S_1$, $S_2$, ($S_3$), the Ernst potential diverges.
 The horizons $\mathcal H_1$, $\mathcal H_2$ are marked as black lines on the
 $\zW$-axis.
 \newline
 Parameters:
 $\phi=\frac{3}{4}\pi$, $w=10/3$, $w'=0.3$ (left panel) and
 $\phi=-0.1$, $w=1.3$, $w'=0.5$ (right panel).}
 \label{fig:1}
\end{figure}

\begin{figure}
 \vspace{1ex}
 \includegraphics[width=\textwidth]{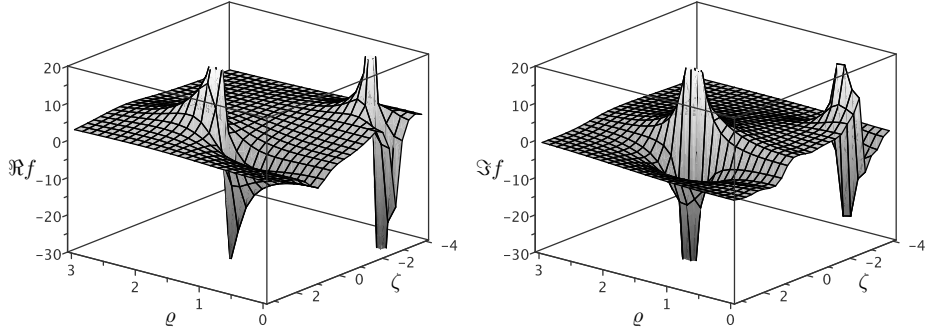}
 \caption{Singular Ernst-Potential: Real and imaginary part of
 the Ernst potential $f$ for the example configuration in
 Fig.~\ref{fig:1}, right panel.} 
 \label{fig:2}
\end{figure}

To tackle this problem we ask whether 
the determinant in the denominator of the representation
\eqref{Yamazaki} of the Ernst potential $f$ has zeros off the axis.
A numerical study for a large number of parameter values
shows that this is indeed the case, see Figs.~\ref{fig:1} and
\ref{fig:2}. As a consequence, the Ernst potential becomes
singular at these zeros (it can be shown that the numerator does not
vanish at the same coordinate positions), i.e. 
there exist
\emph{singular rings} outside the horizons. Our numerical
investigations seem to indicate
that \emph{every double-Kerr-NUT solution with the special parameter relations
\eqref{alpha1} suffers from the presence
of singular rings}.

%%%%%%%%%%%%%%%%%%%%%%%%%%%%%%%%%%%%%%%%%%%%%%%%%%%%%%%%%%%%%%%%%%%%%%%%%%%
\section{Summary}

The stationary equilibrium of two aligned rotating black holes can be 
described by a boundary value problem for two separate (Killing-) 
horizons (see Fig.~\ref{fig:0}).
Applying the inverse (scattering) method, one 
can show that the solution of the problem is a (particular case of the) 
double-Kerr-NUT solution (a solution originally generated by a two-fold 
B\"acklund transformation of Minkowski space). The regularity 
conditions  to be satisfied by the  metric on the axis of symmetry  
outside the two  horizons  restrict   the number of  free  parameters  
entering the solution. The resulting 3-parameter solution  (written in 
dimensionless  coordinates) does not satisfy the characteristic 
condition $8\pi|J| < A$ for each of the two black holes ($J$: angular 
momentum, $A$: area of the horizon). Since this inequality is a consequence 
of the geometry of trapped surfaces in the interior vicinity of the 
event horizon of every sub-extremal black hole, there exists no 
stationary equilibrium configuration for two aligned sub-extremal black 
holes.

%%%%%%%%%%%%%%%%%%%%%%%%%%%%%%%%%%%%%%%%%%%%%%%%%%%%%%%%%%%%%%%%%%%%%%%%%%%

\begin{acknowledgements}
We would like to thank Marcus Ansorg
and David Petroff
for many valuable discussions.
This work was supported by the Deutsche
For\-schungsgemeinschaft (DFG) through the
Collaborative Research Centre SFB/TR7
``Gravitational wave astronomy''.
\end{acknowledgements}

%%%%%%%%%%%%%%%%%%%%%%%%%%%%%%%%%%%%%%%%%%%%%%%%%%%%%%%%%%%%%%%%%%%%%%%%%%%

\end{document}